\documentclass[preprint,5p]{elsarticle}
\usepackage{graphicx}
\usepackage{dcolumn}
\usepackage{bm}  
\usepackage{amssymb}  
\usepackage{hyperref}

\hypersetup{colorlinks=true, urlcolor=blue, citecolor=blue}
\usepackage[displaymath]{lineno}
\usepackage{natbib}
\biboptions{sort&compress}

\title{\vspace{-15mm}\fontsize{18pt}{10pt}\selectfont\textbf{Characteristics of 
fast timing MCP-PMTs in magnetic fields}}

\author[anl,odu]{Mohammad~Hattawy}
\author[anl]{Junqi~Xie\corref{cor}}
\author[bnl]{Mickey~Chiu}
\author[anl]{Marcel~Demarteau}
\author[anl]{Kawtar~Hafidi}
\author[anl]{Edward~May}
\author[anl]{Jos\'e~Repond}
\author[anl]{Robert~Wagner}
\author[anl]{Lei~Xia}
\author[jlab]{Carl~Zorn}

\address[anl]{Argonne National Laboratory, 9700 S. Cass Ave., Argonne, IL 60439, USA}
\address[odu]{Old Dominion University, 5115 Hampton Blvd., Norfolk, VA 23529, USA}
\address[bnl]{Brookhaven National Laboratory, 2 Center St., Upton, NY 11973, USA} 
\address[jlab]{Thomas Jefferson National Accelerator Facility, 12000 Jefferson Ave., Newport News, VA 23606, USA}
\cortext[cor]{Corresponding author: jxie@anl.gov (J. Xie)}

\date{\today}

\begin{document}

\begin{abstract}

 The motivation of this paper is to explore the 
   parameters that affect the performance of Microchannel Plate Photomultiplier Tubes (MCP-PMTs) in magnetic 
   fields  with the goal to guide their design to achieve a high magnetic field tolerance. 
MCP-PMTs based on two different designs were tested.
 The magnetic field tolerance 
   of MCP-PMT based on a design providing independently biased voltages showed a significant 
   improvement (up to 0.7 T) compared to the one utilizing an internal resistor 
   chain design (up to 0.1 T), indicating the importance of individually adjustable voltages.
The effects of the rotation angle of the MCP-PMT
   relative to the magnetic field direction and of the bias voltage between the 
   photocathode and the top MCP were extensively investigated using the MCP-PMT based on the independently 
   biased voltage design.
It was found that the signal amplitude of the MCP-PMT exhibits an enhanced 
   performance at a tilt angle of $\pm$8$^{\circ}$, due to the 8$^{\circ}$ bias angle of the MCP 
   pores.
The maximum signal amplitude was observed at different bias voltages depending on the magnetic field strength.
\end{abstract}

\maketitle


\section{Introduction} \label{sec:level1}
The Electron-Ion Collider (EIC) \cite{1}, which is recommended in the 2015 Long 
Range Plan for Nuclear Science \cite{2} as the highest priority for a new 
facility construction in the US, aims to revolutionize our understanding both 
of nucleon and nuclear structure and of nuclear dynamics in the many-body 
regime, where strongly coupled relativistic quantum fluctuations and 
non-perturbative effects combine to give a dynamical origin to nuclear mass and 
spin. The broad physics program of the EIC requires a large multipurpose 
spectrometer able to measure a plethora of physics processes over a wide range
of energies and solid angles. 
Particular to the EIC is the requirement of particle indentification, 
i.e., the separation of electrons, pions, kaons, and protons (e/$\pi$/K/p) in the final 
state in processes such as semi-inclusive deep inelastic scattering and charm production.

To address the broad physics potential of the EIC, 
several detector concepts are being proposed, including the BeAST 
\cite{3} and the sPHENIX  \cite{4} concepts from 
Brookhaven National Laboratory (BNL), the JLEIC full acceptance detector 
\cite{5} from Thomas Jefferson National Accelerator Facility (JLab), and the 
TOPSiDE 5D particle flow detector \cite{6} from Argonne National Laboratory 
(ANL).  
These detector concepts feature different layouts of 
sub-systems, which have been worked out to varying detail.  
Common to all concepts are the use of  
Time-Of-Flight (TOF) systems and imaging Cherenkov detectors for hadron 
particle identification. 
Integration of these sub-systems into the central 
detector requires to placing their photo-sensors in the non-uniform fringe field 
of the solenoidal magnet.
Thus particle identification at the EIC requires low-cost photon sensors with picosecond timing 
resolution, millimeter spatial resolution, high rate capability, and last but not least
high radiation and magnetic field tolerance.

The microchannel plate photomultiplier tube (MCP-PMT) \cite{7} is a compact 
photosensor consisting of a photocathode for photon-electron conversion, two 
MCPs in a stacked chevron configuration for electron amplification and a 
readout system for charge collection. 
The compact design and confined electron 
amplification by secondary electron emission inside the micron size MCP pores 
provide the MCP-PMT with picosecond timing resolution and millimeter position 
resolution, ideal for application in time-of-flight systems and imaging Cherenkov detectors.  
The LAPPD collaboration \cite{8} between universities, U.S.  national 
laboratories, and industrial partners developed the technology to 
manufacture the world’s largest MCP based photosensor, the Large-Area 
Picosecond Photon Detector (LAPPD$^{TM}$). A critical aspect of the LAPPD$^{TM}$ 
technology is its use of low-cost, very large area (20 $\times$ 20 cm$^2$) MCPs \cite{9} 
within an all glass vacuum envelope. 
The MCPs used in LAPPDs$^{TM}$ are made from bundled and fused capillaries of borosilicate glass 
functionalized through atomic-layer deposition \cite{10,11,12} of 
conductive and secondary-electron emissive material layers.
This revolutionary process eliminates the chemical etching and hydrogen firing steps 
employed in traditional MCP manufacturing, which caused the glass to become brittle 
and resulted in strong ion feedback.
These features and the inherent mechanical stability of 
borosilicate glass allows the production of exceptionally large area MCPs 
with long lifetime \cite{13} and low background noise rates \cite{14}. 

As integral part of the LAPPD project, a dedicated fabrication facility 
\cite{15} capable of producing 6$\times$6~cm$^2$ MCP-PMTs based on the LAPPD design was 
built at Argonne National Laboratory.
The facility served as intermediate production facility while preparing for mass production with our
industrial partner, Incom, Inc \cite{16}. 
To date the Argonne facility produced several dozens of 6$\times$6~cm$^2$ MCP-PMTs which
were provided to various users for early evaluation.
As Incom, Inc.  cranks up mass production of LAPPDs$^{TM}$, the Argonne fabrication 
facility will be converted into an R\&D platform for LAPPD$^{TM}$ design
optimizations geared to specific applications. 
Within a fast turn around, small size (6$\times$6~cm$^2$) MCP-PMTs 
based on different designs can be produced and can be tested either on the test bench or in particle beams.
Once availble the optimized design can be transferred directly to 
Incom, Inc. for LAPPD$^{TM}$ mass production. 

In this paper, we report on tests in magentic fields of two 6$\times$6~cm$^2$ MCP-PMTs based on 
different designs, as produced in the Argonne fabrication facility.
We describe the different designs in section 2, the magnetic field tolerance measurement 
setup in section 3, while the experimental results are presented and discussed in 
section 4; conclusions are drawn at the end of the paper.

\begin{figure*}[tbp]
\centering \includegraphics[scale=1.1]{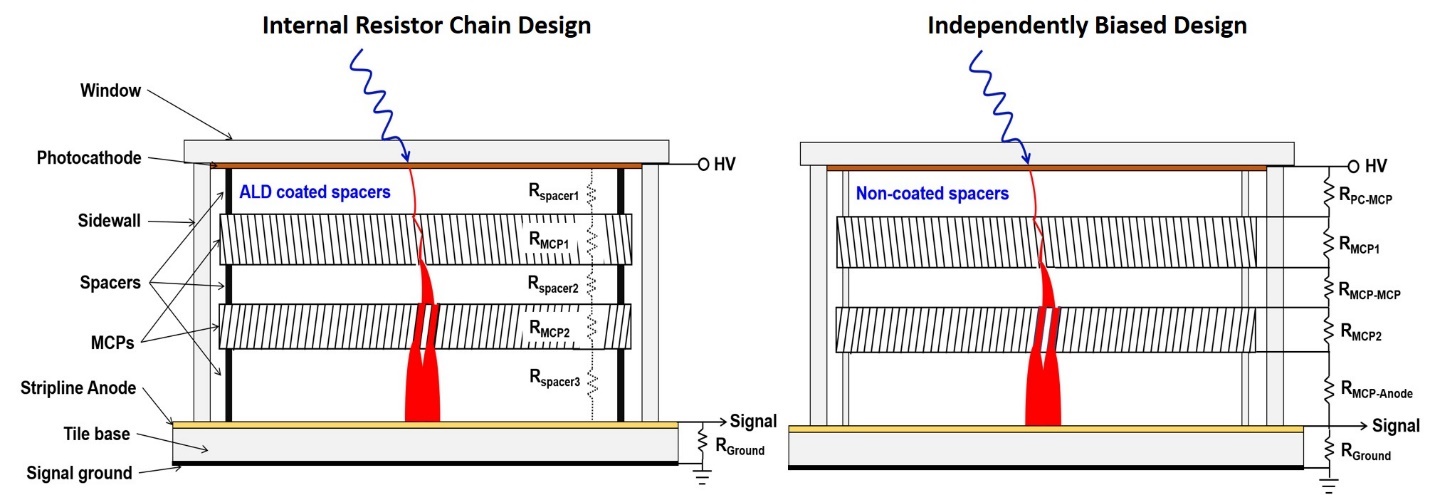}
\caption{Schematic diagrams of the internal resistor chain design (left) and 
   the independently biased design (right). The equivalent electrical circuit 
   in the internal resistor chain design is noted as dashed line connections.  
   Notice the major difference of using ALD coated spacers (resistors) in the 
   internal resistor chain design and non-coated spacers (insulators) in the
   independently biased design.} \label{fig:design}
\end{figure*}

\section{Designs of the MCP photodetector} \label{sec_design}
Two MCP-PMTs based on different designs were tested in this study: the internal 
resistor chain design and the independently biased design. The former relies on 
ALD coated MCPs and spacers inside the MCP-PMT for bias voltage distribution, 
while the latter relies on an external high voltage divider for bias voltage 
distribution.  

\subsection{Internal resistor chain MCP-PMT design} \label{}
The internal resistor chain MCP-PMT design is adapted from the original 
LAPPD$^{TM}$ design \cite{17}. The left panel of Figure~\ref{fig:design} shows 
a schematic of the internal resistor chain MCP-PMT design. The sealed vacuum 
package consists of a photocathode, two MCPs, three grid spacers and a 
stripline anode.  An air-sensitive alkali antimonide photocathode is deposited 
on the inside surface of the top glass window, and the electronic connection is 
provided by a pre-coated nichrome layer at the edges of the top window to apply 
high voltage.  Two MCPs with pores of 8$^{\circ}$ bias angles are placed in chevron 
geometry to prevent drift of positive ions to the photocathode and to ensure a 
well-defined first strike of the incoming photoelectrons. The MCPs used here 
are sliced from the same ALD coated 20 $\times$ 20 cm$^2$ MCPs used
for LAPPD$^{TM}$ production, 
featuring a pore size of 20 $\mu$m, a length to diameter (L:d) ratio of 60:1 
and an open to full area ratio of 65\%.  Glass spacers are used between the 
photocathode and the top MCP, between the MCPs, and between the bottom MCP and 
the anode to separate individual components and support the stack 
configuration. The stripline anode is made through silk-screening of silver 
strips onto the glass tile base, and each stripline is grounded through a resistor.  
It is important to note that the MCPs and glass spacers are all coated with 
resistive materials via the ALD method, making the whole detector stack an internal 
resistor chain, as indicated by the dashed line circuit in 
Figure~\ref{fig:design}. When a single high voltage (HV) is applied to the 
photocathode, the applied HV is distributed between the internal components, 
controlled by the resistances of the ALD coated MCPs and glass spacers. Signals 
generated by incident photons are picked up from the stripline anodes and 
routed to an oscilloscope or an electronic waveform digitizer. 

The internal resistor chain design only requires one HV connection from the outside to the 
inside of the vacuum as provided by the pre-coated nichrome mask on the 
top window. This simple design offers the advantage of ease of 
implementation and potentially low production cost. However, processing and testing of the 
fabricated MCP-PMTs reveal several drawbacks: (a) the HV distribution 
relies on the resistance ratios between the spacers and MCPs, where it is challenging 
to identify precisely matched resistances for the MCPs and spacers; (b) the 
fabrication of MCP-PMT requires thorough baking and scrubbing of the MCPs under 
vacuum for outgassing, while it has been shown that the resistances 
of ALD coated MCPs and spacers are reduced unevenly during this 
process, possibly resulting in mismatched resistances of MCPs and spacers; (c) 
once the detector is sealed, there is no way to individually optimize the MCP’s 
performance as the bias voltage on each MCP cannot be adjusted individually; (d) 
the absolute quantum efficiency (QE) of the photocathode cannot be measured 
using the traditional method as the photocurrent (nA level) generated from 
incident photons is dwarfed by the continuous bias current ($\mu$A level) of 
the resistor chain.      

\begin{figure*}[tbp]
\centering \includegraphics[scale=0.7]{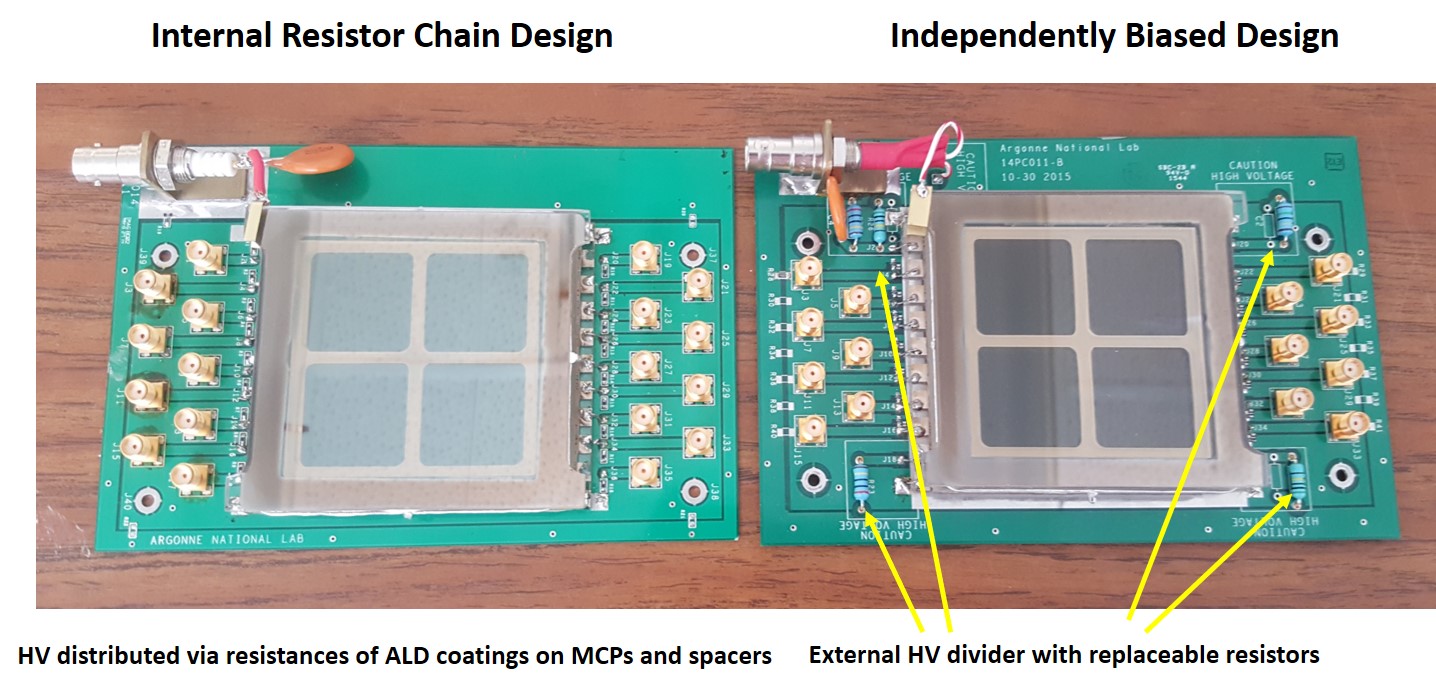}
 \caption{Pictures of MCP-PMTs with the internal resistor chain design (left) 
   and independently biased design (right). Simple readout circuit boards were 
   designed to hold the MCP-PMTs. Note that an external HV divider with 
   replaceable resistors was integrated into the readout board of the 
   independently biased MCP-PMT design. } \label{fig:pics}
\end{figure*}

\subsection{Independently biased MCP-PMT design} \label{sec_design}
The independently biased MCP-PMT design (IBD) offers the option to optimize the
performance of each MCP individually.
 A schematic of the new IBD configuration is shown in the right panel 
of Figure~\ref{fig:design}. The major differences compared to the 
previously described internal resistor design include: (1) 
the spacers are bare glass grids with no ALD coating on the surface, so the 
spacers can be treated as insulators; (2) ultra-thin stainless steel shims with 
the same pattern as grid spacers are attached between the spacers and the MCP 
surfaces to provide HV connections; (3) finger tabs are implemented on each shim, leading 
to the nearest silkscreen printed silver strip contact at one corner, which in turn
provides the HV connection to the outside.
Four shims are inserted between the upper and lower surfaces of the two MCPs.
The new IBD design is based on a minimal modification of the internal resistor chain design, using 
shims and corner strip lines for the HV connection, while no pins are required to provide 
high voltage on the MCPs and in the gaps. Figure~\ref{fig:pics} shows a  
photograph of a sealed MCP-PMT based on the independently biased design (right). 
Simple readout circuit boards were designed and fabricated to hold the MCP-PMTs. An external resistor 
chain HV divider is integrated into the readout board of the independently biased 
MCP-PMT design so that only one HV source is necessary. The 
bias voltage of individual MCPs can be independently adjusted by altering the values of the 
corresponding resistors. 

\begin{figure*}[tbp]
\centering \includegraphics[scale=0.6]{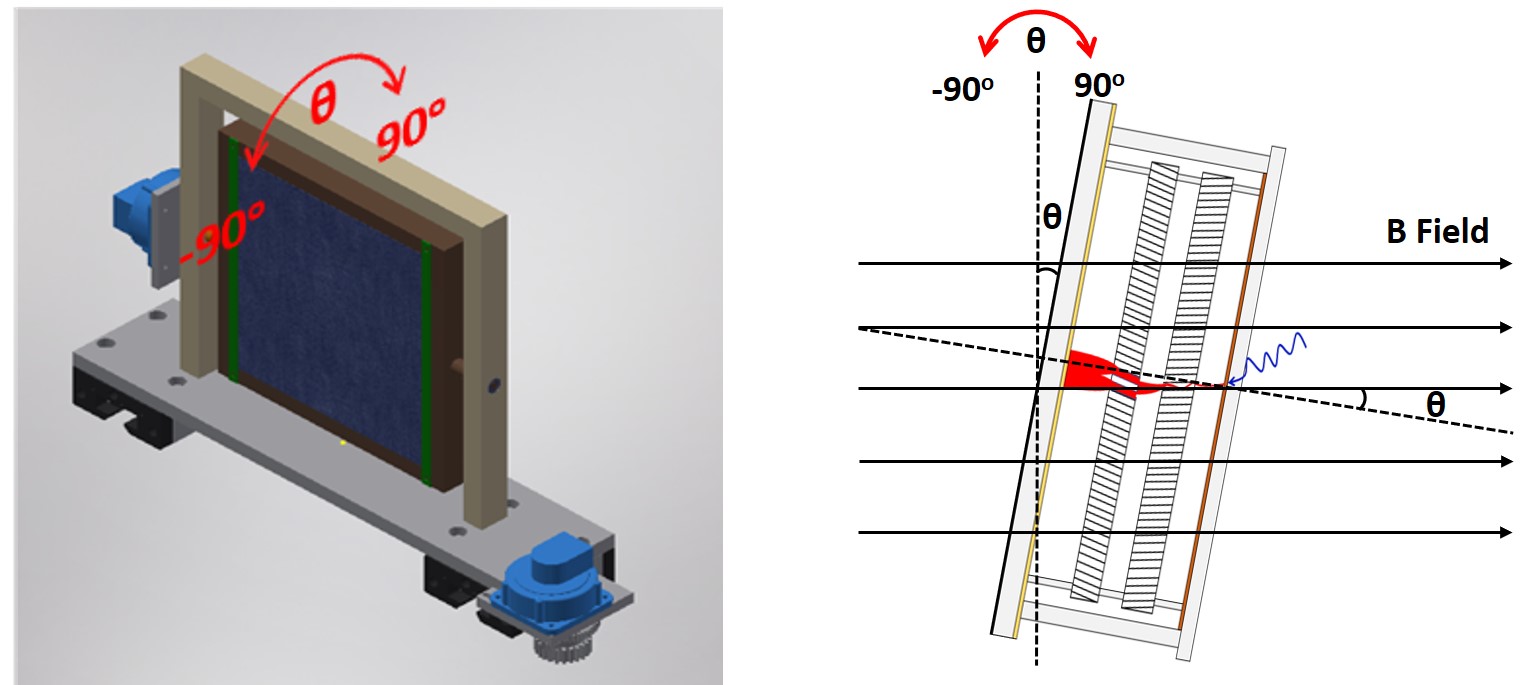}
\caption{(left) AutoCAD drawing of the custom designed magnetic field tolerance 
   test platform. The central part is rotatable with an angle 
   -90$^{\circ}~\leq~\theta~\leq~$90$^{\circ}$.  (right) Schematic of the 
  setup of the MCP-PMT rotated by an angle $\theta$ relative to the 
   magnetic field direction. } \label{fig:theta}
\end{figure*}

\section{Magnetic field tolerance test facility} \label{sc}
Argonne National Laboratory acquired a decommissioned superconducting magnet from
a magnetic resonance imaging (MRI) scanner.
 The primary goal of this magnet is to perform the precise calibration
of the various magnetic probes for the g-2 muon experiment \cite{18}. The MRI magnet provides a 
large bore with a diameter of 68 cm and a very homogeneous field (7 ppb/cm), 
with a tunable magnetic field strength of up to 4 Tesla.  We assembled a characterization 
system compatible with the solenoid magnet to test the performance of the 
6$\times$6~cm$^2$ MCP-PMTs in strong magnetic fields of this magnet. A 
non-magnetic, light-tight dark box was built to 
contain the MCP-PMTs during their tests. The dark box was held on a platform with the 
detector surface normal to the direction of the magnetic field. The position of 
the dark box was adjusted so that the center of the MCP photodetector was 
aligned with the center of the solenoid magnet. A rotation mechanism was 
integrated into the system, able to rotate the MCP-PMTs by an 
angle $\theta$~(-90$^{\circ}~\leq~\theta~\leq~$~90$^{\circ}$), as illustrated  in 
Figure~\ref{fig:theta}.

\begin{figure}[tbp]
\centering \includegraphics[scale=0.4]{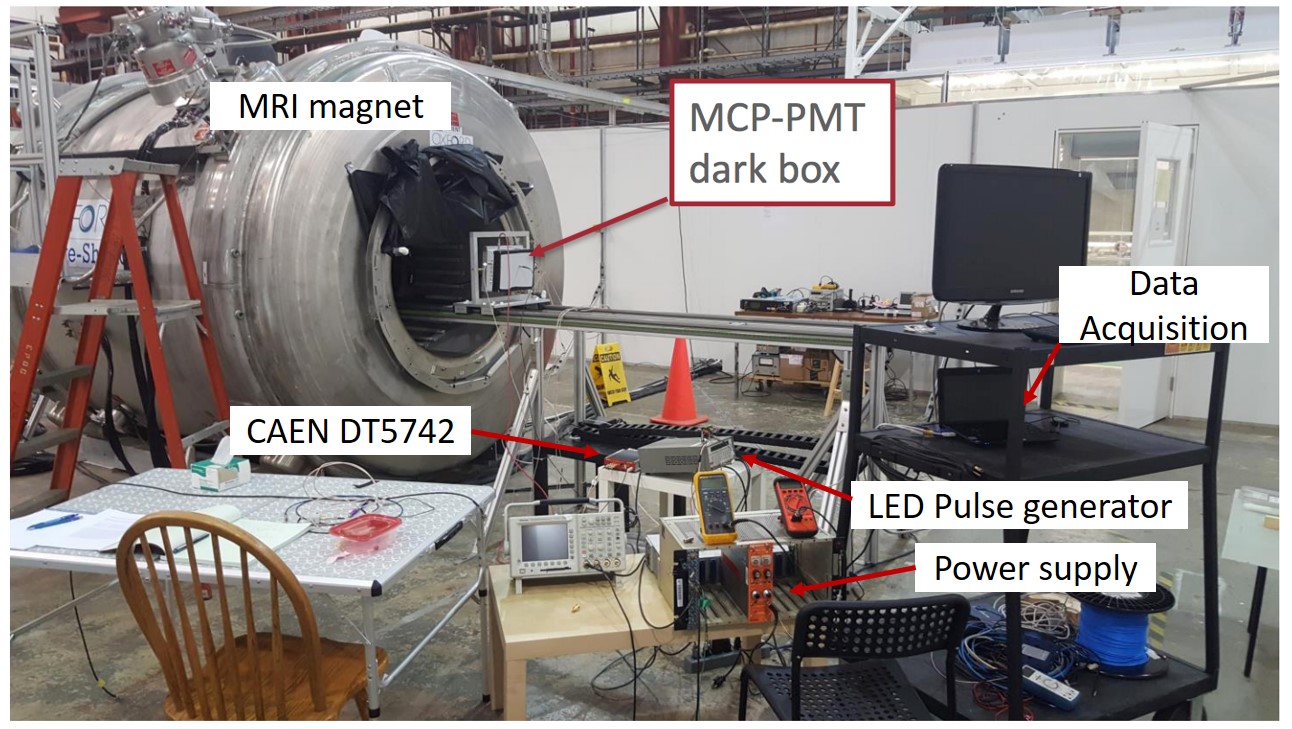}
\caption{Photograph of the magnetic field tolerance testing system.} \label{fig:4}
\end{figure}

\begin{figure*}[tbp]
\centering \includegraphics[scale=0.7]{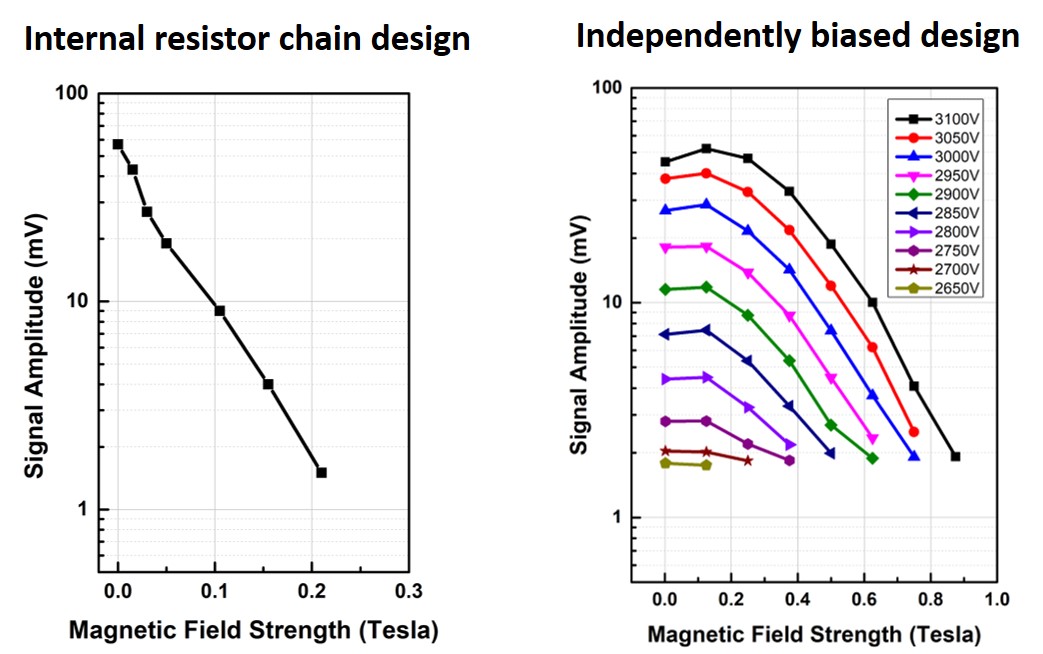}
\caption{Signal amplitude versus magnetic 
   field strength: the internal resistor chain design (left) and the independently 
   biased design (right).
   } \label{fig:5}
\end{figure*}

Figure~\ref{fig:4} shows a picture of the entire magnetic field tolerance testing 
system.  A 405 nm light-emitting diode (LED) driven by a pulse generator provided
the light source. The light was guided into the dark box via an 
optical fiber.  High voltage was applied to the MCP-PMTs from a power supply 
with continuous voltage control. Signals collected at the striplines were read 
out through a DT5742 desktop digitizer \cite{19} with a sampling rate of 5 GS/s, as
produced by CAEN (Costruzioni Apparecchiature Elettroniche Nucleari S.p.A.).  
The digitizer is based on a switched capacitor array of DRS4 (Domino Ring 
Sampler) chips \cite{20} and features 16 analog input channels, and one additional analog 
input for a fast trigger. 

A similar MRI magnet with tunable magnetic field up to 3 Tesla and a similar platform but without the rotation
mechanism were available for MCP-PMT testing at the University of Virginia. 
The following measurements of the MCP-PMT based on the internal resistor chain 
design were performed at University of Virginia, while
measurements of the MCP-PMT based on the independently biased design were performed at Argonne.

\section{Results and discussion} \label{}
The operational principle of MCP-PMTs relies on the electron multiplication process 
where the MCP pore walls are bombarded with secondary electrons. Each 
pore of the MCP has an internal diameter of 
20~$\mu$m with the inner wall processed with resistive and secondary emissive 
coating layers, which act as an independent electron multiplier. When the MCP-PMT is 
operated in a magnetic field, the trajectories of electrons during the electron 
multiplication process are affected by the Lorentz force due to the 
presence of both electric and magnetic fields. We studied the MCP-PMT performance 
as a function of magnetic field strength, rotation angle, and photocathode to MCP electric 
field strength. 

\subsection{Dependence on the magnetic field strength} \label{}
The performance of MCP-PMTs based on the above two designs
was tested in the magnetic field  at a zero rotation angle $\theta$, i.e., where the direction of the 
magnetic field is normal to the surface of the MCP photodetector.
A 405 nm pulsed LED with a fixed light intensity corresponding to 10 photoelectrons
was used as light source.
The signal amplitude versus magnetic field strength is shown in Fig.~\ref{fig:5}. 

\begin{figure}[tbp]
\centering \includegraphics[scale=0.65]{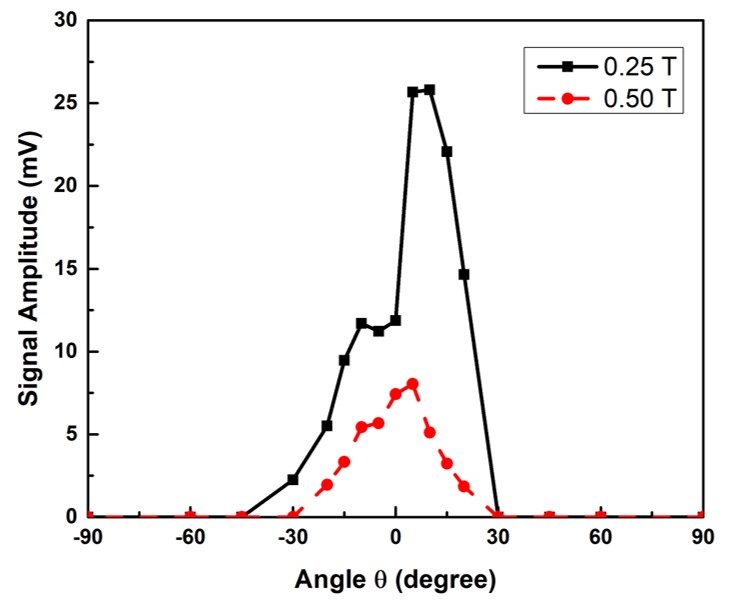}
\caption{The response of the MCP-PMT as a function of the tilt angle $\theta$ 
   between the normal to the MCP-PMT window and the direction of the magnetic 
   field.  The two peaks around -8$^{\circ}$ and 8$^{\circ}$ are related to the 
   bias angle of the MCP pores. Note that the 
   intensities of the two peaks are not the same due to the different effect 
   of the top and bottom MCPs. } 
   \label{fig:6}
\end{figure}

The MCP-PMT based on the internal resistor chain design shows a poor magnetic 
field tolerance, the signal amplitude drops by a factor of 6 when the magnetic 
field increases from 0 to 0.1 Tesla, and another factor of 6 when the 
field increases to 0.2 Tesla. This rapid decrease is mainly due to the resistances 
of the MCPs and spacers having been significantly changed during the baking and 
scrubbing process, resulting in a bias voltage mismatch of the two MCPs. 
These results indicate that the MCP-PMT based on the 
resistor chain design is not suitable for applications in magnetic fields over 0.1 T. 

On the other hand, the MCP-PMT based on the independently biased design shows 
a significantly improved tolerance to magnetic fields. The performance was measured
at various magnetic field strengths and applying various bias high voltages. 
At a fixed magnetic field strength, the 
signal amplitude increases with increasing bias high voltage. 
This behavior confirms our previous measurements of MCP-PMTs in a
negligible magnetic field \cite{21}. At a fixed bias voltage of 3100 V, 
the signal amplitude of the MCP-PMT increases slightly as the magnetic field 
strength increases to 0.2~T, and then is seen to decrease as the magnetic field strength 
continues to increase, and eventually is reduced below 5 
mV at a magnetic field strength of 0.7~T. With lower bias voltages, the signal amplitudes are
seen to be reduced already at lower field strength. 
As these results show, the decrease in signal strength can to some extend be compensated by
increased bias voltages.

\subsection{Dependence on the tilt angle} \label{}
The signal strength as function of tilt angle $\theta$  between the normal to the MCP-PMT window and the direction of the 
magnetic field, as shown in Figure~\ref{fig:theta}, was investigated using the MCP-PMT based on the independently 
biased design.
We applied a fixed high 
voltage of 3000 V and rotated the tilt angles $\theta$ from 
-90$^{\circ}$ to 90$^{\circ}$.  
Figure~\ref{fig:6} shows the signal amplitude as a function of the tilt angle $\theta$ for two magnetic 
field strengths of 0.25 and 0.5 Tesla, respectively.  The signal amplitude 
shows a strong angle dependence with vanishing signals outside the range of -30$^{\circ}~\leq~\theta~\leq~$30$^{\circ}$ 
and two maxima at $\pm$8$^{\circ}$. 
The latter are related to the 8$^{\circ}$ bias angle of the MCP pores 
and their chevron configuration.  When the direction of one MCP pore is aligned 
with the direction of the magnetic field, the MCP-PMT shows an enhanced 
magnetic field tolerance. The signal maximum at +(-)8$^{\circ}$ corresponds to the 
position where the direction of the pores of the top (bottom) MCP is aligned with the direction of 
the magnetic field.

\begin{figure}[tbp]
\centering \includegraphics[scale=0.32]{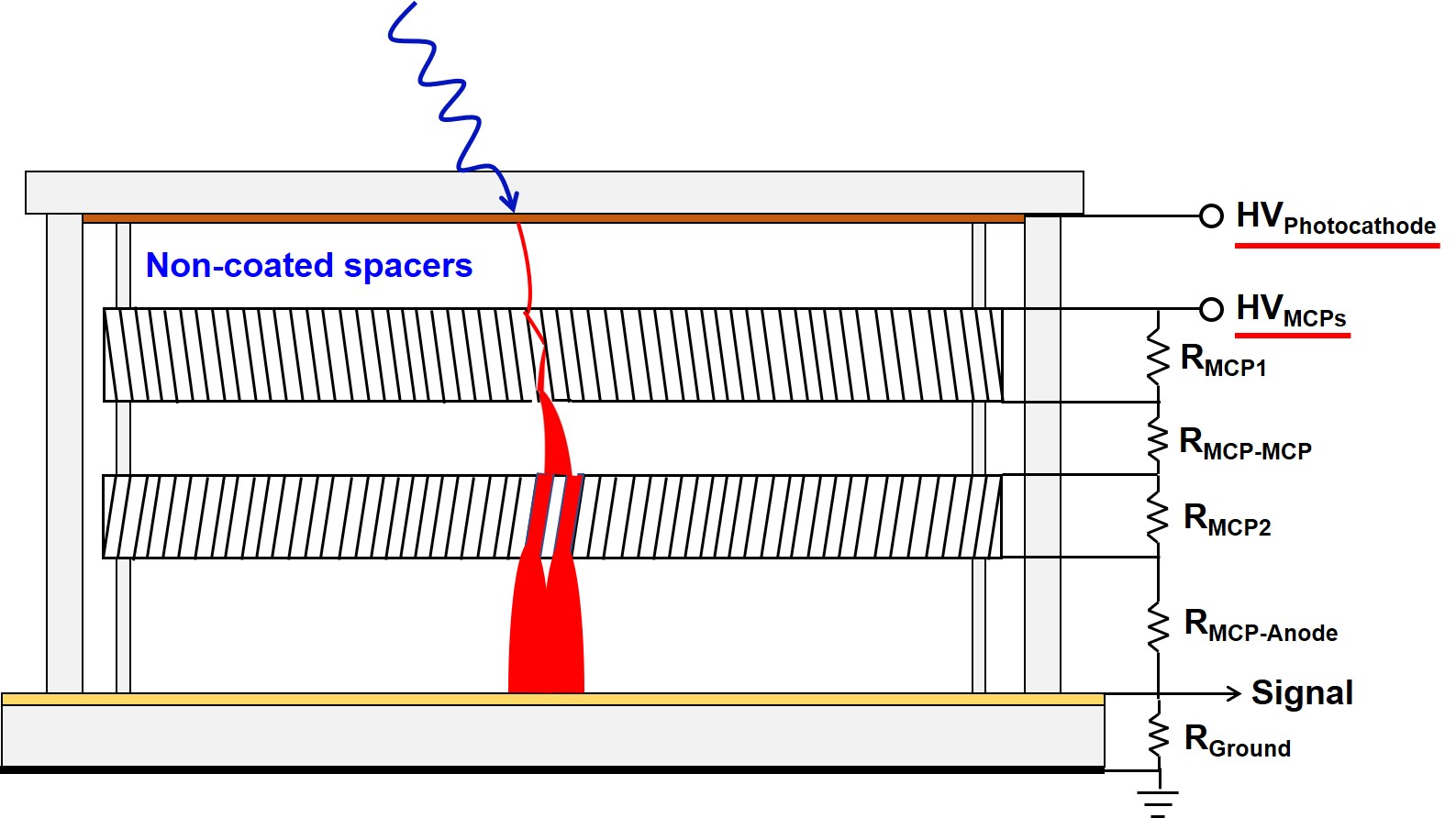}
\caption{The electrical circuit of HV connections devised specifially to be able to vary the gap voltage 
   between the photocathode and the top MCP.} \label{fig:7}
\end{figure}

\subsection{Dependence on the gap high voltage} \label{}
The MCP-PMTsignal amplitude as a function of applied HV to the gap between the 
photocathode and the top MCP was studied at different magnetic field 
strengths. Figure~\ref{fig:7} shows the circuit diagram which allowed to vary the applied gap 
voltage. While the HV$_{MCPs}$ 
was kept at a fixed value, the HV$_{Photocathode}$ was varied to adjust the gap voltage. 

\begin{figure}[tbp]
\centering \includegraphics[scale=0.6]{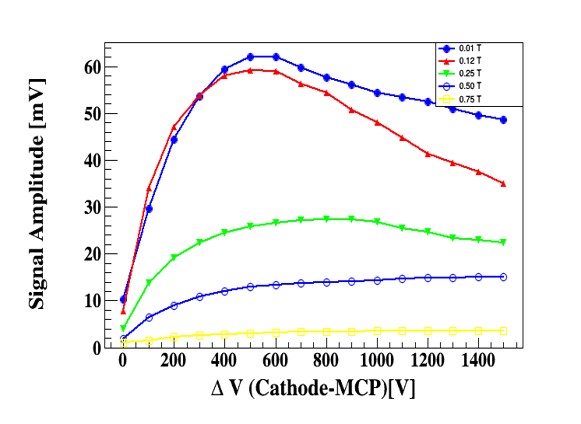}
\caption{Performance of the MCP-PMTs in terms of signal amplitude as a function 
   of gap voltage applied between the photocathode and top MCP in different 
   magnetic fields.} \label{fig:8}
\end{figure} 

Figure~\ref{fig:8} shows the signal amplitude versus gap voltage for a selection of magnetic 
field strengths. With low magnetic fields, the signal 
amplitude increases as the gap voltage increases and reaches a maximum at a gap 
voltage of $\sim$ 500 V. 
Further increasing the gap voltage beyond 500 V results in a decreasing signal amplitude.
 This effect is related to the energy of the primary electrons, as studied previously \cite{22}. 
As these studies showed, the yield of secondary emissions of ALD coated materials is highest
when the primary electron energy is around 300 – 500 eV, corresponding to a gap HV at $\sim$ 500 V.
With higher primary electron energies, the electrons penetrate deeper into the coating, thus  
again reducing the yield of secondary emission electrons.
At high magnetic fields, 
the magnetic field strength becomes the main parameter affecting the secondary 
emission process. 
The secondary yield is not seen to 
decrease anymore with primary electron energy over 500 eV, resulting in a 
continuously increasing signal amplitude with increasing gap voltages.

\section{Conclusions}
Two 6$\times$6 cm$^2$ MCP-PMTs based on the internal resistor chain design and 
the independently biased design were fabricated at Argonne National Laboratory and 
characterized in magnetic fields. The behavior of 
the MCP-PMT signal amplitude was investigated as a function of the magnetic 
field strength, the distribution of bias voltage, the tilt angle, and the gap 
voltage. It was found that the MCP-PMT based on the internal resistor chain design 
shows a magnetic field tolerance only up to 0.1~T. With the independently biased voltage 
design, the magnetic field tolerance of the MCP-PMT is significantly improved,
up to 0.7~T. 
These findings indicate the importance of ensuring that both MCPs are operated at their 
optimal bias 
voltages when operated in high magnetic fields. As the magnetic field 
strength increases, the signal amplitude of the MCP-PMT decreases while being operated 
at a constant bias voltage.
However, the reduction of signal amplitude can be compensated 
by increasing the operation voltage, extending the range of operability in 
high magnetic fields. Due to the original MCP bias angle of 8$^{\circ}$ and 
the chevron configuration, the pores of both MCPs can not be aligned simultaneously with the direction of 
the magnetic field. The MCP-PMT shows higher signal amplitudes when either 
MCP pores is aligned with the direction of the magnetic field, where the 
direction of the top MCP pores exhibits a stronger impact on the signal amplitude.
Increasing the bias voltage applied on the gap between the photocathode and the top MCP results in 
a maximum signal amplitude with gap voltage around 500 V at low magnetic 
fields, while a continuously increasing signal amplitude is observed at high magnetic 
fields. 

\section{Acknowledgments}
The authors thank Frank Skrzecz (Engineer at ANL) for his mechanical 
engineering support; Joe Gregar (Scientific Glass Blower at ANL) for his 
talented work on glass parts; Mark Williams and Wilson Miller (Professors at 
University of Virginia) for their arrangement of the University of Virginia MRI 
magnet usage; Peter Winter (Physicist at ANL) for his arrangement of the 
Argonne 4-Tesla magnetic facility usage; and many people from the LAPPD 
collaboration for their advice and assistance. This material is based upon 
work supported by the U.S. Department of Energy, Office of Science, Office of 
High Energy Physics, Office of Nuclear Physics, under contract number 
DE-AC02-06CH11357. Work at Thomas Jefferson National Accelerator Facility was 
supported by the U. S. Department of Energy, Office of Science under contract 
No. DE-AC05-06OR23177. Work at Brookhaven National Laboratory was supported by 
the U. S. Department of Energy, Office of Science under contract No.  
DE-SC0012704. This work was also partially supported by the EIC R\&D funding 
from the Office of Nuclear Physics and Office of Science of the U.S.  
Department of Energy.

\end{document}